\documentclass[9pt,twocolumn,twoside]{opticajnl}
\journal{opticajournal} 

\setboolean{shortarticle}{true}

\usepackage{lineno}

\title{Modal formulation of Kirchhoff's law for reciprocal structures}
\author[1]{D. Tihon \thanks{denis.tihon@uclouvain.be}} 
\author[2]{S. Withington \thanks{stafford.withington@physics.ox.ac.uk}} 
\affil[1]{\footnotesize ICTEAM Institute, Universit\'{e} catholique de Louvain, Belgium}
\affil[2]{\footnotesize Department of Physics, University of Oxford, Oxford, UK}

\usepackage{amsmath}
\usepackage{amssymb}
\usepackage{esint}
\usepackage{xcolor}
\usepackage{bm}

\newcommand{\vect}[1]{\bm{#1}}

\newcommand{\mat}[1]{\underline{\underline{#1}}}

\newcommand{\matn}{\mat{\spatialscal{n}}}
\newcommand{\Pabs}{P_\text{abs}}

\newcommand{\rpos}{{\vect{r}}}
\newcommand{\rposp}{{\vect{r}'}}
\newcommand{\psipos}{{\bm{\psi} }}
\newcommand{\psiposp}{{\bm{\psi}'}}
\newcommand{\kt}{\vect{k}_t}
\newcommand{\ktp}{\vect{k}'_t}

\newcommand{\spatial}[1]{\overrightarrow{\vect{#1}}}
\newcommand{\spatialscal}[1]{\overrightarrow{#1}}
\newcommand{\spectral}[1]{\tilde{\vect{#1}}}
\newcommand{\spectralscal}[1]{\tilde{#1}}
\newcommand{\angular}[1]{\hat{\vect{#1}}}
\newcommand{\angularscal}[1]{\hat{#1}}
\usepackage[normalem]{ulem}

\begin{abstract}
Kirchhoff's law provides a relation between the incident fields a structure can absorb and those it thermally emits. It is generally formulated as a relation between the angle-dependent emissivity and absorptivity of a structure. In this letter, we propose to extend the definition of the absorptivity and emissivity to account for interference and coherence effects. These new definitions are used to derive a modal form of Kirchhoff's law for reciprocal structures, in which ``absorptive" and ``emissive" modes can be paired. We show that the formulation of Kirchhoff's law strongly depends on the basis used to express the fields. Three different formulations are proposed, which are related to three popular bases: plane waves characterized by their direction of propagation, plane waves characterized by their transverse wave vector, and the spatial distribution of the tangential electric and magnetic fields along a closed surface that contains the structure.
\end{abstract}

\setboolean{displaycopyright}{false} 

\begin{document}
\maketitle

\section{Introduction}
For a long time, it has been known that there exists a fundamental relation between the electromagnetic fields a given structure can absorb and those it thermally emits. Historically, Kirchhoff formulated this relation as a universal ratio between the frequency-dependent absorptivity of a material and its frequency-dependent emissivity. Following this pioneering work, several refinements to the law have been proposed to include the angular or polarization dependence of the absorptivity~\cite{Ohwada_1988, Greffet_2018}, account for complex geometries~\cite{Wojszvzyk_2019} or extend the relation to luminescent emitters~\cite{Band_1988, Greffet_2018} and non-reciprocal structures~\cite{Ohwada_1988, Snyder_1998, Greffet_2018, Guo_2022}. 

To date, the most common formulations use a plane wave basis to describe the incident and emitted fields. The absorptivity is defined as the angle-dependent absorption cross-section of the structure when it is illuminated by an incident plane wave~\cite{Greffet_2018}. The emissivity is defined as the angle-dependent emission cross-section of the structure, i.e. the cross-section required for a black body to emit the same power in the direction considered. Those formulations intrinsically suffer from two major drawbacks. First, They cannot handle evanescent power transfer nor thermal generation of stored reactive energy, intrinsically limiting their scope. Second, they only deal with part of the information available since those formulations do not account for interference effects that may appear when the structure is illuminated by a coherent superposition of plane waves, nor do they account for the partial coherence of the spontaneously emitted fields. To circumvent these limitations, two different approaches have been proposed, which we will refer to as the \emph{current-based} and \emph{modal} approaches. 
 
 In the current-based approach~\cite{Rytov_1989, Tihon_2021}, the absorptivity is defined as the power absorbed by the structure when it is illuminated by a pair of electric and/or magnetic dipoles. The emissivity is defined as the cross-spectral power density tensor of the partially coherent fields, i.e. the first-order correlation tensor of the fields. The power transferred from a single electric or magnetic dipole to the structure is related to the intensity of the electric or magnetic fields emitted by the structure at the position of the dipole. The amplitude of the interference effect between two dipoles is related to the correlation of the fields emitted at the locations of the dipoles. This formulation can handle reactive and partially coherent fields, solving the two aforementioned limitations. However, it is based on Lorentz reciprocity theorem and is only valid for reciprocal structures.

The second approach has been proposed by Miller and co-authors~\cite{Miller_2017} and is based on a modal decomposition of the fields that can be absorbed or emitted by a given structure. The authors postulate the existence of a particular basis in which the coupling between the different modes vanishes. Based on thermodynamic arguments, general laws between incident and emitted modes are proposed for reciprocal and non-reciprocal structures. While providing an intuitive formulation and being applicable to non-reciprocal structures, this approach still suffers from some limitations. First, the incident and emitted modes are defined using a physical ground, but no mathematical definition is proposed in the paper. While providing a high-level understanding of fundamental relations between emission and absorption, it is hardly applicable to practical situations, such as the prediction of the thermal fields emitted at a pair of locations from a set of measurements in absorption. Moreover, these modes are normalized with respect to the active power they carry, a definition that is not compatible with reactive fields.

In this letter, we use the current-based approach to derive a modal formulation of reciprocity that is compatible with near field coupling. We show that the modal formulation is not unique and strongly depends on the definition used for the ``incident" fields or on the basis used to express the fields. We derive three different modal Kirchhoff's laws related to three different commonly used bases: a spatial description of the fields, a spatial-spectral description of the fields crossing a plane (i.e. in the momentum space) and a propagating plane wave spectrum characterized by the angle of propagation. In the rest of this letter, these bases will be referred to as the \emph{spatial}, \emph{spectral} and \emph{angular} bases, respectively. We show that the latter basis, which corresponds to the basis traditionally used in the literature, is a particular case that yields an intuitive formulation of Kirchhoff's law. For clarity and brevity, only the main results are presented in this Letter. Detailed calculations are available in the Supplementary Materials.

\begin{figure}
    \centering
    \includegraphics[width = 8cm]{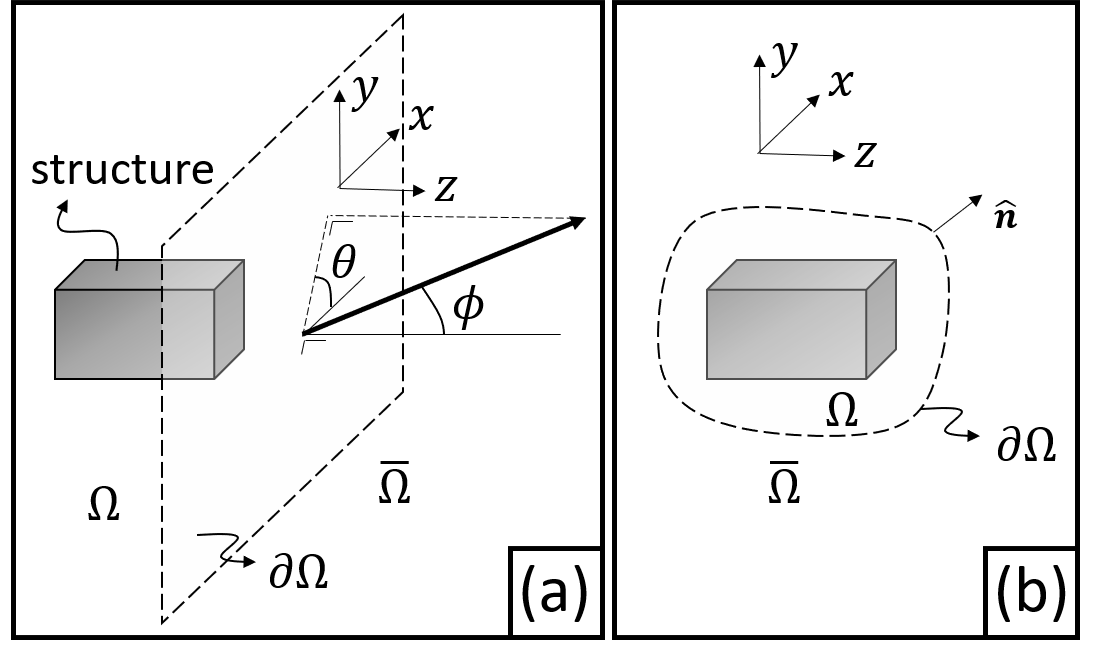}
    \caption{Geometry and conventions used to formulate the modal version of Kirchhoff's law (a) using a plane wave basis and (b) using a spatial description of the fields.}
    \label{fig:geometry}
\end{figure}

We start by considering the classical approach, where the absorptivity and emissivity of a structure are expressed as angle-dependent quantities that describe the coupling to incident and emitted polarized plane waves. The geometry studied is illustrated in Fig. \ref{fig:geometry}(a). The space is split into two halves: one half $\Omega$ that contains the structure, and the other half $\bar{\Omega}$ that is empty (i.e. free-space). Between the two halves is a plane interface $\partial \Omega$ on which the incident and emitted fields are sampled. When the structure acts as an emitter, we consider the partially coherent fields it emits in the empty half-space $\bar{\Omega}$. When it acts as an absorber, we consider the power it absorbs when it is illuminated by sources located in $\bar{\Omega}$. In the following, the ``incident" fields are defined as the fields that would be generated by the sources in the absence of the structure (i.e. if $\Omega$ was empty).

The structure is reciprocal. It is also linear and time-invariant, so that the analysis can be done easily in the spectral domain using phasors. An $\exp(j \omega t)$ time-dependence of the fields and currents is implicitly assumed. Using the equivalence principle~\cite{Guissard_2003}, the structure can be characterized by considering only the fields crossing the interface $\partial \Omega$. We denote these fields at any position $\rpos=(x,y,z)$ along the interface as a column vector $\spatial{F}(\rpos) = [\spatial{E}(\rpos); \eta_0 \spatial{H}(\rpos)]$, with $\eta_0$ the free-space impedance. Note that, in the following, hats, tildes and arrows will be used to denote quantities expressed in the angular, spectral and spatial bases, respectively .

If the interface is sufficiently far away from the structure or the sources, the evanescent spectrum can be neglected. The fields along the interface are decomposed into plane waves propagating toward directions $\psipos = (\theta, \phi)$, with $\theta$ the azimuthal angle ($\theta \in [0, 2\pi]$) and $\phi$ the elevation angle ($\phi \in [0, \pi]$). We define the plane wave spectrum using the angle-dependent column vector $\angular{F}(\psipos)$, whose entries correspond to the amplitude of the TE and TM plane waves defined such that:
\begin{equation}
\label{eq:paper01}
\begin{split}
    \spatial{F}(\rpos) &= \dfrac{1}{4\pi} \int_{\phi=0}^{\pi} \int_{\theta = 0}^{2\pi} \mat{\angularscal{d}}(\psipos) \cdot \angular{F}(\psipos) \\
    & \hspace{-1cm} \times 
    \exp\Big(-j k_0 \sin(\phi) \big( \cos(\theta) x + \sin(\theta) y\big) \Big) \sin(\phi) \, d\theta \, d\phi,
\end{split}
\end{equation}
with $k_0$ the free-space wavenumber and $\mat{\angularscal{d}}$ the matrix containing the directions of the two polarizations:
\begin{subequations}
\label{eq:paper03}
\begin{gather}
    \mat{\angularscal{d}}(\psipos) = 
    \begin{bmatrix}
        \angular{e}(\psipos) & \angular{m}(\psipos) \\
        \angular{m}(\psipos) & - \angular{e}(\psipos)
    \end{bmatrix}, 
    \\
    \angular{e}(\psipos) = \begin{bmatrix}
        -\sin(\theta) \\ \cos(\theta) \\ 0
    \end{bmatrix},
    \hspace{1cm}
    \angular{m}(\psipos) = \begin{bmatrix}
        \cos(\theta)\cos(\phi) \\ \sin(\theta)\cos(\phi) \\ \sin(\phi)
    \end{bmatrix}.
\end{gather}
\end{subequations}

As explained by Withington \textit{et al.}~\cite{Stafford_2007}, any linear time-invariant absorber can be characterized using a second-order tensor on which the incident fields are left- and right-projected. This tensor is part of a more general quantity describing the interaction between the absorber and any generalized force~\cite{Stafford_2017}. Diagonal elements of this tensor correspond to the power absorbed by the structure when it is illuminated by a single source. Off-diagonal elements describe the interference effects that will take place if two or more phase-locked sources are illuminating the device simultaneously. This tensor is a generalization of the concept of \emph{mixed losses} originally introduced by Rytov~\cite{Rytov_1989}. Using the plane wave convention described above, this quantity $\mat{\angularscal{C}}_\text{abs}$ is defined such that the power $\Pabs$ absorbed by the structure when it is illuminated by incident fields $\angular{F}^\text{in}$ reads
\begin{equation}
\label{eq:paper04}
    \Pabs = \iint \iint \angular{F}^{\text{in},\dagger}(\psipos) \cdot \mat{\angularscal{C}}_\text{abs}(\psipos, \psiposp) \cdot \angular{F}^\text{in}(\psiposp) \, d\psipos \, d\psiposp,
\end{equation}
with $d\psi = \sin(\phi) \, d\theta d\phi / 4\pi$ and $\vect{A}^\dagger$ the conjugate transpose of $\vect{A}$. Given the definition of the incident fields, it can be noticed that this definition of $\mat{\angularscal{C}}_\text{abs}$ is ambiguous. Indeed, no plane wave with $\phi<\pi/2$ can be generated by sources located in $\bar{\Omega}$. For simplicity, given that $\bar{\Omega}$ is empty so that no plane wave emitted at the interface can be reflected back toward the structure, we consider that $\mat{\angularscal{C}}_\text{abs} = 0$ when $\phi<\pi/2$ or $\phi'<\pi/2$.

The ``raw" mixed losses can be difficult to interpret or manipulate. For this reason, it was proposed by Withington \textit{et al.}~\cite{Stafford_2007} to reformulate it using a modal expansion:
\begin{equation}
\label{eq:paper06}
    \mat{\angularscal{C}}_\text{abs}(\psipos, \psiposp) = \sum_i \angularscal{\Lambda}_{i}^\text{abs} \angular{F}^\text{abs}_i (\psipos) \angular{F}^{\text{abs}, \dagger}_i (\psiposp).
\end{equation}
Each mode of the expansion corresponds to an independent degree of freedom through which the structure absorbs power. The values $\Lambda_i$ describe the amount of power dissipated by each mode and the vectors $\angular{F}^\text{abs}_i$ are related to the incident fields distribution to which the structure is sensitive. Due to the properties of the mixed losses of finite structures (Hermitian and Hilbert-Schmidt), it is possible to find a decomposition made of orthogonal vectors $\angular{F}_i$.

We now look at the structure as a thermal or luminescent emitter. Fields will be spontaneously emitted by random fluctuations taking place inside the structure. These random fluctuations are due to the finite temperature of the structure. In some situations, it is possible to increase the rate of these transitions, a phenomenon known as luminescence. In those two cases, the fields originate from random events, and must thus be described statistically. They are said to be partially coherent. 
One way to describe these fields is to use a first-order correlation function, the so called \emph{cross-spectral power density tensor}~\cite{Tervo_2004}. In the spatial domain, this tensor is defined as
\begin{equation}
    \mat{\spatialscal{C}}(\rpos, \rposp) = \langle \spatial{F}(\rpos) \spatial{F}^\dagger(\rposp) \rangle,
\end{equation}
where the angular brackets denote an ensemble average and $\vect{A}\vect{B}^\dagger$ denotes the tensor product of $\vect{A}$ with $\vect{B}^\dagger$, so that entry $(i,j)$ of the resulting matrix corresponds to product of the $i^\text{th}$ entry of A and the complex conjugate of the $j^\text{th}$ entry of $\vect{B}$. Note that, for statistically stationary processes, the ergodic assumption can be made to replace the ensemble average by a time average. 

Diagonal elements of this tensor correspond to the average intensity of the fields at a given location, parallel to a given direction. Off-diagonal entries describe the correlation of the fields at different positions or along different directions. This tensor has many interesting properties, such as the fact that it can be propagated using Maxwell's equations~\cite{Wolf_1955}. Thus, one can determine the correlation of the fields at any pair of locations in the empty half-space $\bar{\Omega}$ from its value along the interface $\partial \Omega$. 

The cross-spectral power density tensor can be expressed using the plane-wave basis described above. Then, the partially coherent fields emitted by the structure read
\begin{equation}
\label{eq:paper10}
    \mat{\angularscal{C}}_\text{em}(\psipos, \psiposp) = \langle \angular{F}^\text{em}(\psipos) \angular{F}^{\text{em},\dagger}(\psiposp) \rangle,
\end{equation}
with $\angular{F}^\text{em}$ the angular spectrum emitted by the structure during one ``experiment", which is averaged over an ensemble of identical ``experiments". The cross-spectral power density tensor is Hermitian, semi-definite positive and Hilbert-Schmidt. Thus, it admits a modal decomposition~\cite{Tervo_2004}:
\begin{equation}
\label{eq:paper07}
    \mat{\angularscal{C}}_\text{em}(\psipos, \psiposp) = \sum_i \Lambda_i^\text{em} \angular{F}^\text{em}_i(\psipos) \angular{F}_i^{\text{em}, \dagger}(\psiposp),
\end{equation}
where each mode corresponds to a fully coherent field distribution, and different modes are superimposed incoherently.

Obviously, $\mat{\angularscal{C}}_\text{abs}(\psipos, \psiposp)$ corresponds to a generalization of the absorptivity of the structure, the latter corresponding to the diagonal elements of the former. A similar observation can be made by comparing $\mat{\angularscal{C}}_\text{em}(\psipos, \psiposp)$ with the emissivity. As we may expect from Kirchhoff's law, the two quantities are related by the relation (cf. Supplementary Materials)
\begin{equation}
    \label{eq:paper02}
    \mat{\angularscal{C}}_\text{em}(\psipos, \psiposp) = \angularscal{C}_\text{Th} \, \mat{\angularscal{C}}_\text{abs}^*(\bar\psipos, \bar\psiposp),
\end{equation}
with $\angularscal{C}_\text{Th}$ a constant that is independent of the angular coordinates and $\bar\psipos = (\pi + \theta, \pi-\phi)$ the angle pointing to the direction opposite to $\psipos$. The detailed expression of $\angularscal{C}_\text{Th}$ can be found in the Supplementary Materials. 

\eqref{eq:paper02} can be straightforwardly extended to a modal formulation using \eqref{eq:paper06}:
\begin{align}
    \mat{\angularscal{C}}_\text{abs}(\psipos, \psiposp) 
        &= \sum_i \Lambda_i \angular{F}^\text{abs}_i(\psipos) \angular{F}^{\text{abs},\dagger}_i(\psiposp)
        \nonumber
        \\
        & \hspace{-1.5cm} \rightarrow \hspace{0.5cm}
    \mat{\angularscal{C}}_\text{em}(\psipos, \psiposp) 
        = \sum_i \Lambda_i \angularscal{C}_\text{Th} \angular{F}_i^{\text{abs},*}(\bar\psipos) \angular{F}_i^{\text{abs},T}(\bar\psiposp).
        \label{eq:paper17}
\end{align}
The modal expansion tells us that, for each independent degree of freedom through which a structure can absorb energy, there is an equivalent degree of freedom through which it can emit energy. Moreover, for a given mode, to the combination of incident plane waves to which the structure is sensitive corresponds a similar combination of counter-propagating planes waves emitted with the same polarization (linear, right- or left-handed circular, etc.). Any retardation effect in the absorption is related to an earliness effect in the emission. 

The angular spectrum is handy to study far-field phenomena. However, to study near-field coupling and evanescent waves, one needs to consider complex angles, resulting in a cumbersome and non-intuitive formulation. In this case, it is generally easier to characterize the plane waves through their transverse wave vector $\kt = (k_x, k_y)$. In that case, the definition of the amplitude of the TE and TM waves should be slightly modified. First, to one $\kt$ are associated four different plane waves: the TE and TM waves crossing the reference plane toward or away from the structure. Thus, we define the vector $\spectral{A} = [\spectralscal{A}_e^-; \spectralscal{A}_e^+; \spectralscal{A}_m^-; \spectralscal{A}_m^+]$ corresponding to the concatenation of the amplitudes related to the TE (subscript $e$) and TM (subscript $m$) plane waves emitted toward the structure (superscript $-$) or away from it (superscript $+$). The amplitudes are then defined such that, on the interface between the two half-spaces, 
\begin{equation}
\label{eq:paper14}
\begin{split}
    \spatial{F}(\rpos) &= \dfrac{1}{4\pi^2} \int_{k_x = -\infty}^{\infty} \int_{k_y = -\infty}^{\infty} \mat{\spectralscal{d}}(\kt) \cdot \spectral{F}(\kt) 
    \\ 
    & \hspace{2cm}
    \times \exp\Big(-j (k_x x + k_y y) \Big) \, dk_x \, dk_y,
\end{split}
\end{equation}
with $\mat{\spectralscal{d}}$ the matrix containing the directions associated with each polarization and transverse wave vector:
\begin{equation}
\label{eq:paper13}
    \mat{\spectralscal{d}}(\kt) = 
    \begin{bmatrix}
        \spectral{e}^-(\kt) & \spectral{e}^+(\kt) &
            \spectral{m}^-(\kt) & \spectral{m}^+(\kt)  \\
        \spectral{m}^-(\kt) & \spectral{m}^+(\kt) &
            - \spectral{e}^-(\kt) & - \spectral{e}^+(\kt)
    \end{bmatrix}, 
\end{equation}
where 
\begin{equation}
    \spectral{e}^\pm(\kt) = \dfrac{1}{k_t}\begin{bmatrix}
        -k_y \\ k_x \\ 0
    \end{bmatrix},
    \hspace{0.5cm}
    \spectral{m}^\pm(\kt) = - \dfrac{1}{k_0 k_t}\begin{bmatrix}
        k_x k_z^\pm \\ k_y  k_z^\pm \\ k_t^2
    \end{bmatrix}.
\end{equation}
with $k_t = \sqrt{k_x^2+k_y^2}$ and $k_z^\pm = \pm \sqrt{k_0^2-k_t^2}$ if $k_0>k_t$ and $k_z^\pm = \mp j \sqrt{k_t^2-k_0^2}$ otherwise.

Using this slightly modified basis, the mixed losses can be defined as the second-order tensor that satisfies
\begin{equation}
\label{eq:paper08}
    \Pabs = \iint_{\kt} \iint_{\ktp} \spectral{F}^{\text{in}, \dagger}(\kt) \cdot \mat{\spectralscal{C}}_\text{abs}(\kt, \ktp) \cdot \spectral{F}^{\text{in}}(\ktp) \, d\kt \, d\ktp,
\end{equation}
with $d\kt = dk_x \, dk_y/4 \pi^2$. Similarly, the cross-spectral power density tensor is defined as
\begin{equation}
\label{eq:paper11}
    \mat{\spectralscal{C}}_\text{em}(\kt, \ktp) = \langle \spectral{F}^\text{em}(\kt) \spectral{F}^{\text{em},\dagger}(\ktp) \rangle.
\end{equation}

Using these definitions, Kirchhoff's law can be reformulated as (cf. Supplementary Materials):
\begin{equation}
\label{eq:paper05}
    \mat{\spectralscal{C}}_\text{em}(\kt, \ktp) = \dfrac{\spectralscal{C}_\text{Th}}{k_z^{+} \big({k'}_z^{+}\big)^*} \mat{P} \cdot \mat{\spectralscal{C}}_\text{abs}^*(-\kt, -\ktp) \cdot \mat{P},
\end{equation}
with $\spectralscal{C}_\text{Th}$ a constant that is independent from the spectral coordinate, whose detailed value is provided in the Supplementary Materials, and $\mat{P}$ a permutation matrix defined as
\begin{equation}
    \mat{P} = \begin{bmatrix}
        0 & 1 & 0 & 0 \\
        1 & 0 & 0 & 0 \\
        0 & 0 & 0 & -1 \\
        0 & 0 & -1 & 0
    \end{bmatrix}.
\end{equation}
Using the modal decomposition of the mixed losses, one obtains the modal formulation of Kirchhoff's law in the spectral basis:
\begin{align}
    \mat{\spectralscal{C}}_\text{abs}(\kt, \ktp) 
        &= \sum_i \Lambda_i \spectral{F}^\text{abs}_i(\kt) \spectral{F}^{\text{abs},\dagger}_i(\ktp)
        \nonumber
        \\
        & \hspace{-1.5cm}\rightarrow \hspace{0.5cm}
    \mat{\spectralscal{C}}_\text{em}(\kt, \ktp) 
        = \sum_i \Lambda_i \spectralscal{C}_\text{Th} \bigg(\dfrac{1}{k_z^+} \mat{P} \cdot \spectral{F}^{\text{abs},*}_i(-\kt)\bigg) 
        \nonumber \\
        & \hspace{2.2cm} \times 
        \bigg(\dfrac{1}{{k'}_z^+} \mat{P} \cdot \spectral{F}^{\text{abs},*}_i(-\ktp)\bigg)^\dagger.
\end{align}
The spectral formulation leads to qualitatively similar conclusions as the angular formulation. However, some additional factors $1/k_z^+$ and $1/({k'}_z^+)^*$ appear. These factors naturally emerge from the basis used. A different definition of the basis could result in different factors. A direct consequence of these factors is that a decomposition of the mixed losses into orthogonal modes may not lead to an orthogonal modal decomposition of the cross-spectral power density tensor. Thus, the Mercer's expansion of the mixed losses, which is the decomposition generally used in the literature, may not correspond one-to-one with the Mercer's expansion of the cross-spectral power density tensor. It may seem counter-intuitive but does not raise any mathematical issue, since the modal decomposition of a tensor is not unique.

The spectral formulation is well suited to the study of far-field and near-field effects when a reference plane can be drawn between the structure and the sources that interact with the structure. However, in many practical situations, it is interesting to study the response of a small structure to sources located all around it, or to look at the fields emitted by a structure in any direction of space. One possibility to do so is to consider a finite volume in which the structure is embedded and consider the incident and emitted fields tangential to the surface of the volume (see Fig. \ref{fig:geometry}(b)). To do so, we introduce the $\matn$ tensor, which is defined such that, at any position on the surface $\partial \Omega$, $\matn \cdot [\spatial{E}; \eta_0 \spatial{H}] = [\eta_0 \spatial{n} \times \spatial{H}; \spatial{n} \times \spatial{E}]$. 
Using this definition, the tangential part of the fields $\spatial{F}_t$ can be reformulated as
\begin{equation}
    \spatial{F}_t(\rpos) = - \matn(\rpos) \cdot \matn(\rpos) \cdot \spatial{F}(\rpos).
\end{equation}

In this new basis, the mixed losses are defined such that
\begin{equation}
\label{eq:paper09}
    \Pabs = \oiint_{\rposp \in \partial \Omega} \oiint_{\rpos \in \partial \Omega} \spatial{F}_t^\dagger(\rpos) \cdot \mat{\spatialscal{C}}_\text{abs}^\text{tan}(\rpos, \rposp) \cdot \spatial{F}_t(\rposp) d\rpos \, d\rposp.
\end{equation}
Similarly, the cross-spectral power density tensor is defined as
\begin{equation}
\label{eq:paper12}
    \mat{\spatialscal{C}}_\text{em}^\text{tan}(\rpos, \rposp) = \langle \spatial{F}_t^\text{em}(\rpos) \spatial{F}_t^{\text{em},\dagger}(\rposp) \rangle,
\end{equation}
with $\spatial{F}_t^\text{em}(\rpos)$ the tangential fields thermally generated by the structure at position $\rpos$ of the surface $\partial \Omega$.

A third formulation of Kirchhoff's law can be derived between these two quantities (cf. Supplementary Materials), which reads
\begin{equation}
\label{eq:paper15}
    \mat{\spatialscal{C}}_\text{em}^\text{tan}(\rpos, \rposp) = \spatialscal{C}_\text{Th} \, \matn(\rpos) \cdot \mat{\spatialscal{C}}_\text{abs}^{\text{tan},*}(\rpos, \rposp) \cdot \matn^T(\rposp).
\end{equation}
The corresponding modal formulation is given by
\begin{align}
    \mat{\spatialscal{C}}_\text{abs}^\text{tan}(\rpos, \rposp) 
        &= \sum_i \Lambda_i \spatial{F}^\text{abs}_i(\rpos) \spatial{F}^{\text{abs},\dagger}_i(\rposp)
        \nonumber
        \\
        & \hspace{-1.5cm} \rightarrow \hspace{0.5cm}
    \mat{\spatialscal{C}}_\text{em}^\text{tan}(\rpos, \rposp) 
        = \sum_i \Lambda_i \spatialscal{C}_\text{Th}  \Big(\matn(\rpos) \cdot \spatial{F}_i^{\text{abs},*}(\rpos)\Big) 
        \nonumber \\
        & \hspace{2cm} 
        \times \Big(\matn(\rposp) \cdot \spatial{F}_i^{\text{abs},*}(\rposp)\Big)^\dagger.
\end{align}
The physical interpretation of this formulation of Kirchhoff's law should be formulated with caution, since not any incident field realization is possible. Indeed, $\spatial{F}_i^\text{abs}$ does not correspond to a tangential incident field distribution that is physically realizable, but to the vector on which those physically realizable fields should be projected. Keeping this in mind, the amplitude of the vector on which the magnetic (resp. electric) field should be projected is proportional to the amplitude of the electric (resp. magnetic) field thermally emitted at the same location. Their orientations are orthogonal. 

Obtaining a basis-dependent formulation of Kirchhoff's law may seem counter-intuitive, however it is a direct consequence of the definitions that have been used for the mixed losses and the cross-spectral power density. The mixed losses are defined as a second-order tensor on which the incident fields should be projected to compute the power absorbed by the structure. The power absorbed by the structure for a given illumination should be independent from the basis in which the fields are represented. Thus, it appears clearly from \eqref{eq:paper04}, \eqref{eq:paper08} and \eqref{eq:paper09} that the mixed losses are a covariant quantity. To the contrary, looking at the definition of the cross-spectral power density tensor in \eqref{eq:paper10}, \eqref{eq:paper11} and \eqref{eq:paper12}, the latter is a contravariant quantity. Obviously, the relation between a covariant and a contravariant quantity cannot be independent from the basis used. The intuitive formulation that is obtained in the angular domain might explain why, traditionally, Kirchhoff's law has been developed using a propagating plane wave basis, despite its known limitations. In that respect, current-based formulation~\cite{Rytov_1989, Tihon_2021} might be better suited: as long as the power generated by a current distribution can be expressed as a multiplication of the amplitude of the currents by the amplitude of the fields, the currents behave as covariant quantities. Thus, the associated mixed losses are contravariant.

\section*{Conclusion}
To summarize, traditionally, Kirchhoff's law is formulated as the relation between power absorbed by a structure when it is illuminated by an incident plane wave coming from a given direction and the power the structure spontaneously radiates back in that direction. The former quantity corresponds to the absorptivity of the structure. The latter quantity corresponds to its emissivity. In this letter, we proposed to generalized the concepts of absorptivity and emissivity. The new formulation includes coherence and interference effects and can be translated into different bases. Kirchhoff's law has been reformulated as a mathematical relation between these two quantities. We showed that the absorptivity and emissivity are covariant and contravariant quantities, respectively, so that the formulation of the Kirchhoff's law depends on the basis considered. Three different commonly used bases have been considered: an angular, a spectral and a spatial description of the fields. For each basis, a modal version of the Kirchhoff's law has been proposed. As expected, the traditional Kirchhoff's law is a particular case of the more general laws that have been proposed. 

\section*{Acknowledgements}
Denis Tihon is a Postdoctoral Researcher of the Fond de la Recherche Scientifique - FNRS.

\bibliography{biblio}

\end{document}


\maketitle


\subsection{Current-based formulation of Kirchhoff's law}
Consider a linear time-invariant structure illuminated by electric and magnetic currents $\spatial{K}(\rpos) = [\spatial{J}(\rpos); \spatial{M}(\rpos)/\eta_0]$. It is possible to define the mixed losses $\mat{\spatialscal{C}}_\text{abs}^{K}$ such that the power absorbed by the structure reads:
\begin{equation}
    \Pabs = \iiint \iiint \spatial{K}^\dagger(\rpos) \cdot \mat{\spatialscal{C}}_\text{abs}^{K}(\rpos, \rposp) \cdot \spatial{K}(\rposp) \, d\rpos \, d\rposp
\end{equation}
Provided that the structure is reciprocal and its temperature $T$ is uniform, it can be proven that the cross-spectral power density tensor of the thermal fields generated by the structure reads~\cite{Tihon_2021}
\begin{equation}
\label{eq:appendix03}
    \mat{\spatialscal{C}}_\text{em}(\rpos, \rposp) = 4 \Theta(T) \mat{S} \cdot \mat{\spatialscal{C}}_\text{abs}^{K,*}(\rpos, \rposp) \cdot \mat{S}
\end{equation}
with $C^*$ the complex conjugate of $C$, $\mat{S}$ a sign-changing matrix whose diagonal elements are $[1, 1, 1, -1, -1, -1]$ and whose non-diagonal element are zeros. $\Theta(T)$ is the temperature-dependent Planck's function and reads:
\begin{equation}
    \Theta(T) = \dfrac{\hbar \omega}{\exp\Big(\dfrac{\hbar \omega}{k_B T}\Big) -1}
\end{equation}
with $k_B$ the Boltzmann constant and $\hbar$ the reduced Planck constant. Note that the formulation can be adapted to account for luminescence by introducing a chemical potential, so that the current-based reciprocity can also be used to model luminescent emitters~\cite{Greffet_2018}.


\subsection{Proof of angular Kirchhoff's law}
The proof of the angular version of Kirchhoff's law (cf. Eq. (8) and Eq. (9)) can be decomposed into four steps.
\begin{enumerate}
    \item Compute the plane waves spectrum generated by a given current distribution along the interface.
    \item Postulating that the mixed losses are known in the angular domain, compute the current-based mixed losses by looking at the power dissipated for any current distribution.
    \item Using the current-based formulation of Kirchhoff's law, predict the value of the cross-spectral power density tensor along the interface. 
    \item From the value of the fields along the interface, evaluate the corresponding plane-wave spectrum.
\end{enumerate}

Note that, during the procedure, it will be required to go from the spatial to the angular domain, and vice-versa. Thus, we need to define the inverse of the transform defined in Eq. (1). Since not all the fields realizations are physically possible, the inverse transform is not unique. One possibility is:
\begin{align}
\label{eq:appendix05}
    \angular{F}(\psipos) &= \dfrac{k_0^2 |\cos(\phi)|}{\pi} \iint \dinvangular(\psipos) \cdot \spatial{F}(\rpos) \exp\Big( j k_0 \sin(\phi)\big( \cos(\theta) x + \sin(\theta) y\big)\Big) \, dx \, dy
\end{align}
with $\dinvangular$ a pseudo-inverse of $\mat{\angularscal{d}}$ given by:
\begin{equation}
    \dinvangular = \dfrac{1}{2}\begin{bmatrix}
        -\sin(\theta)
            & \cos(\theta)
            & 0 
            & \dfrac{\cos(\theta)}{\cos(\phi)}
            & \dfrac{\sin(\theta)}{\cos(\phi)}
            & 0 \\
        \dfrac{\cos(\theta)}{\cos(\phi)} 
            & \dfrac{\sin(\theta)}{\cos(\phi)}
            & 0
            & \sin(\theta)
            & -\cos(\theta)
            & 0
    \end{bmatrix}
\end{equation}
This pseudo-inverse has been chosen such that $\dinvangular(\theta, \phi) \cdot \mat{\angularscal{d}}(\theta, \phi) = \mat{1}$ and $\dinvangular(\theta, \phi) \cdot \mat{\angularscal{d}}(\theta, \pi-\phi) = \mat{0}$, with $\mat{1}$ and $\mat{0}$ the unitary and null matrices.

\subsubsection{Incident fields generated by a given current distribution}
The incident fields are defined as the fields that would be generated by the sources if the volume containing the structure was empty. By hypothesis, the other half-space is empty. Thus, the computation of the fields radiated by currents on the interface can be computed in two steps. First, compute the fields generated using the free-space Green's function, then compute the angular spectrum this field distribution corresponds to.

Consider a current distribution $\spatial{K}(\rpos) = [\spatial{J}(\rpos); \spatial{M}(\rpos)/\eta_0]$ along the interface between the two half-spaces, with $\spatial{J}$ and $\spatial{M}$ the electric and magnetic currents. Maxwell's equations being linear with respect to the source terms, the current distribution can be decomposed into sheets of current, each sheet being characterized by a given transverse wave vector $\kt$:
\begin{equation}
\label{eq:appendix06}
    \spectral{K}(\kt) = \iint \spatial{K}(\rposp) \exp\big(j \kt \cdot \rposp) d\rposp
\end{equation}
The field generated at any position $\rpos$ along the interface then reads~\cite{Tihon_PW}
\begin{equation}
\label{eq:appendix07}
    \spatial{F}(\rpos) = - \dfrac{k_0 \eta_0}{8\pi^2} \iint_{\kt} \dfrac{1}{k_z^+}  \mat{\spectralscal{d}}(\kt) \cdot \mat{\spectralscal{d}}^T(\kt) \cdot \spectral{K}(\kt) \exp(-j \kt \cdot \rpos) \, dk_x \, dk_y
\end{equation}
with $\mat{\spectralscal{d}}$ defined in Eq. (11). 

Combining Equations \eqref{eq:appendix05}, \eqref{eq:appendix06} and \eqref{eq:appendix07}, the angular spectrum generated by the current distribution reads:
\begin{equation}
\label{eq:appendix01}
    \angular{F}^\text{in}(\psipos) = -\dfrac{\eta_0 k_0^2}{2\pi} \mat{\angularscal{d}}^T(\psipos) \cdot \iint \spatial{K}(\rpos) \exp\Big(j k_0 \sin(\phi) \big(\cos(\theta) x + \sin(\theta) y\big)\Big) \, dx \, dy
\end{equation}
with $\mat{d}^T$ the transpose of $\mat{d}$. Note that the following identities have been used:
\begin{align}
    &\int_x \exp\big(j g(k) x\big) dx = 2\pi \delta\big(g(k)\big)
    \\
    &\int_k f(k) \delta\big(g(k)\big) dk = \sum_{k_i | f(k_i) = 0} \dfrac{f(k_i)}{|g'(k_i)|}
\end{align}
with the last sum that is carried out over all the roots $k_i$ of the function $g(k)$ included in the integration domain, and $g'(k)$ the derivative of the function $g$.

\subsubsection{Computation of the current-based mixed losses}
We consider that the angular mixed losses $\mat{\angularscal{C}}_\text{abs}$ of the structure are known. Combining Eq. (3) and \eqref{eq:appendix01}, the total power absorbed reads:
\begin{align}
    \Pabs 
    & = \iint_\psipos \iint_{\psiposp} \angular{F}^{\text{in},\dagger}(\psipos) \cdot \mat{\angularscal{C}}_\text{abs}(\psipos, \psiposp) \cdot \angular{F}^\text{in}(\psiposp) \, d\psipos \, d\psiposp
    \\
    & = \dfrac{\eta_0^2 k_0^4}{(2\pi)^2} \iint_\psipos \iint_{\psiposp} \iint_{\rpos} \iint_{\rposp}
    \spatial{K}^\dagger(\rpos) \cdot \mat{\angularscal{d}}^*(\psipos) \cdot \mat{\angularscal{C}}_\text{abs}(\psipos, \psiposp) \cdot \mat{\angularscal{d}}^T(\psiposp) \cdot \spatial{K}(\rposp) 
     \\
    & \hspace{2cm}
    \times \exp\Big(-j k_0 \sin(\phi) \big(\cos(\theta) x + \sin(\theta) y\big)\Big)
    \nonumber \\
    & \hspace{2cm}
    \times \exp\Big(j k_0 \sin(\phi') \big(\cos(\theta') x' + \sin(\theta') y'\big)\Big) \, d\rpos \, d\rposp \, d\psipos \, d\psiposp \nonumber
    \\
    & \triangleq \iint_\rpos \iint_{\rposp} \spatial{K}^\dagger(\rpos) \cdot \mat{\spatialscal{C}}_\text{abs}^K(\rpos, \rposp) \cdot \spatial{K}(\rposp) \, d\rpos \, d\rposp
\end{align}
with 
\begin{equation}
\label{eq:appendix02}
\begin{split}
    \mat{\spatialscal{C}}_\text{abs}^K(\rpos, \rposp)
    & =
    \dfrac{\eta_0^2 k_0^4}{(2\pi)^2} \iint_\psipos \iint_{\psiposp}
    \exp\Big(-j k_0 \sin(\phi) \big(\cos(\theta) x + \sin(\theta) y\big)\Big) 
    \\ & \hspace{3cm}
    \times \mat{\angularscal{d}}^*(\psipos) 
    \cdot \mat{\angularscal{C}}_\text{abs}(\psipos, \psiposp)
    \cdot \mat{\angularscal{d}}^T(\psiposp)
     \\
    & \hspace{2cm} 
    \times \exp\Big(j k_0 \sin(\phi') \big(\cos(\theta') x' + \sin(\theta') y'\big)\Big) \,  d\psipos \, d\psiposp
\end{split}
\end{equation}
Note that, as defined in the Letter, $d\psipos = \sin(\phi) \, d\theta d\phi / 4\pi$.

\subsubsection{Using current-based Kirchhoff's law}
Using the current-based mixed losses defined in Equation \eqref{eq:appendix02} and combining it with the current-based Kirchhoff's law of Equation \eqref{eq:appendix03}, one obtains: 
\begin{align}
\label{eq:appendix04}
    \mat{\spatialscal{C}}_\text{em}(\rpos, \rposp)
    & =
    \dfrac{\Theta(T) \eta_0^2 k_0^4}{\pi^2} \iint_\psipos \iint_{\psiposp}
    \exp\Big(j k_0 \sin(\phi) \big(\cos(\theta) x + \sin(\theta) y\big)\Big) 
    \\ & \hspace{3cm} \nonumber 
    \times \mat{S} \cdot \mat{\angularscal{d}}(\psipos) 
    \cdot \mat{\angularscal{C}}_\text{abs}^*(\psipos, \psiposp)
    \cdot \mat{\angularscal{d}}^\dagger(\psiposp)
    \cdot \mat{S}
     \\
    & \hspace{2cm}  \nonumber 
    \times \exp\Big(-j k_0 \sin(\phi') \big(\cos(\theta') x' + \sin(\theta') y'\big)\Big) \,  d\psipos \, d\psiposp
\end{align}

\subsubsection{Computation of the angular plane wave spectrum}
The angular transform described in Equation \eqref{eq:appendix05} can be combined with the definition of the cross-spectral power density tensor in the angular domain (Eq. (6)) to obtain the Kirchhoff's law formulated in the angular domain:
\begin{align}
    \mat{\angularscal{C}}_\text{em}(\psipos, \psiposp) 
    & \triangleq \langle \angular{F}^\text{em}(\psipos) \angular{F}^{\text{em},\dagger}(\psiposp) \rangle 
    \\
    &= \Bigg\langle \dfrac{|\cos(\phi)| k_0^2}{\pi} \iint \dinvangular(\psipos) \cdot \spatial{F}^\text{em}(\rpos) \exp\Big( j k_0 \sin(\phi)\big( \cos(\theta) x + \sin(\theta) y\big)\Big) \, d\rpos
    \\ \nonumber
    & \hspace{-1cm} \times \bigg(\dfrac{|\cos(\phi') k_0^2|}{\pi} \iint \dinvangular(\psiposp) \cdot \spatial{F}^\text{em}(\rposp) \exp\Big( j k_0 \sin(\phi')\big( \cos(\theta') x' + \sin(\theta') y'\big)\Big) \, d\rposp \bigg)^\dagger
    \Bigg\rangle
    \\ \label{eq:appendix17}
    &= \dfrac{|\cos(\phi)| |\cos(\phi')| k_0^4}{\pi^2} \iint_\rpos \iint_{\rposp} \dinvangular(\psipos) \cdot 
    \langle \spatial{F}^\text{em}(\rpos) \spatial{F}^{\text{em},\dagger}(\rposp) \rangle 
    \cdot \big(\dinvangular(\psiposp)\big)^\dagger
    \\ \nonumber & \hspace{3cm} \times
    \exp\Big( j k_0 \sin(\phi)\big( \cos(\theta) x + \sin(\theta) y\big)\Big)
    \\ \nonumber & \hspace{3cm} \times
    \exp\Big(-j k_0 \sin(\phi')\big( \cos(\theta') x' + \sin(\theta') y'\big)\Big) \, d\rpos \, d\rposp 
\end{align}
The averaged value that appears in \eqref{eq:appendix17} corresponds to the cross-spectral power density tensor whose value was determined in Equation \eqref{eq:appendix04}. After long but straightforward calculations, it leads to:
\begin{align}
    \mat{\angularscal{C}}_\text{em}(\psipos, \psiposp) &= 16 \, \eta_0^2 \, k_0^{4} \, \Theta(T) \mat{\angularscal{C}}_\text{abs}^*(\bar\psipos, \bar\psiposp)
    \\
    &\triangleq \angularscal{C}_\text{Th} \, \mat{\angularscal{C}}_\text{abs}^*(\bar\psipos, \bar\psiposp)
\end{align}
with $\angularscal{C}_\text{Th} = 16 \, \eta_0^2 \, k_0^{4} \, \Theta(T)$ and $\bar\psipos = (\pi + \theta, \pi-\phi)$, the angle pointing to the direction opposite to $\psipos$.

\subsection{Spectral formulation}
The proof of the spectral formulation is identical to the proof of the angular formulation, except that different transformation rules are used to go from spatial to spectral domain. It can thus be decomposed into four steps:
\begin{enumerate}
    \item Compute the spectral content of the fields generated by a given current distribution.
    \item Postulating that the mixed losses are known in the spectral domain, compute the current-based mixed losses by looking at the power dissipated for any current distribution.
    \item Using the current-based formulation of Kirchhoff's law, predict the value of the cross-spectral power density tensor along the interface.
    \item Evaluate the spectral content of the fields along the interface.
\end{enumerate}

Note that, during the procedure, it will be required to go from the spatial to the spectral domain, and vice-versa. Thus, we need to define the inverse of the transform defined in Eq. (10). Since not all the fields realizations are physically possible, the inverse transform is not unique. One possibility is:
\begin{align}
\label{eq:appendix11}
    \spectral{F}(\kt) &= \iint \dinvspectral(\kt) \cdot \spatial{F}(\rpos) \exp\big( j (k_x x + k_y y)\big) \, dx \, dy
\end{align}
with $\dinvspectral$ a pseudo-inverse of $\mat{\spectralscal{d}}$ given by:
\begin{equation}
    \dinvspectral(\kt) = \dfrac{1}{2}\begin{bmatrix}
        \einvspectral(\kt) & \einvspectral(\kt) & -\minvspectral(\kt) & \minvspectral(\kt) \\
        -\minvspectral(\kt) & \minvspectral(\kt) & -\einvspectral(\kt) & -\einvspectral(\kt)
    \end{bmatrix}^T
\end{equation}
\begin{equation}
    \einvspectral(\kt) = \spectral{e}(\kt)
    , \hspace{2cm}
    \minvspectral(\kt) = - \dfrac{k_0}{k_t k_z^+}\begin{bmatrix}
        k_x \\ k_y \\ 0
    \end{bmatrix}
\end{equation}

\subsubsection{Incident fields generated by a given current distribution}
The incident fields are defined as the fields that would be generated by the sources if the volume containing the structure was empty. By hypothesis, the other half-space is empty. Thus, the computation of the fields radiated by currents on the interface can be computed using the free-space Green's function. 

Consider a current distribution $\spatial{K}(\rpos) = [\spatial{J}(\rpos); \spatial{M}(\rpos)/\eta_0]$ along the interface between the two half-spaces, with $\spatial{J}$ and $\spatial{M}$ the electric and magnetic currents. The plane wave spectrum generated by this current distribution reads~\cite{Tihon_PW}:
\begin{equation}
\label{eq:appendix08}
    \spectral{F}(\kt) = - \dfrac{k_0 \eta_0}{2} \dfrac{1}{k_z^+}  \mat{\spectralscal{d}}^T(\kt) \cdot \iint \spatial{K}(\rpos) \exp\big(j \kt \cdot \rpos) d\rpos
\end{equation}
with $\mat{\spectralscal{d}}$ defined in Eq. (11). 

\subsubsection{Computation of the current-based mixed losses}
We consider that the spectral mixed losses $\mat{\spectralscal{C}}_\text{abs}$ of the structure are known. Combining Eq. (13) and \eqref{eq:appendix08}, the total power absorbed reads:
\begin{align}
    \Pabs 
    & = \iint_{\kt} \iint_{\ktp} \spectral{F}^{\text{in},\dagger}(\kt) \cdot \mat{\spectralscal{C}}_\text{abs}(\kt, \ktp) \cdot \spectral{F}^\text{in}(\ktp) \, d\kt \, d\ktp
    \\
    & = \dfrac{\eta_0^2 k_0^2}{4} \iint_{\kt} \iint_{\ktp} \iint_{\rpos} \iint_{\rposp}
    \dfrac{1}{(k_z^+)^*}\spatial{K}^\dagger(\rpos) \cdot \mat{\spectralscal{d}}^*(\kt) \cdot \mat{\spectralscal{C}}_\text{abs}(\kt, \ktp) \cdot \mat{\spectralscal{d}}^T(\ktp) \cdot \spatial{K}(\rposp) \dfrac{1}{{k'}_z^+}
     \\
    & \hspace{2cm}
    \times \exp\big(-j ( k_x x + k_y y)\big) \, \exp\big(j ( k'_x x' + k'_y y')\big)\, d\rpos \, d\rposp \, d\kt \, d\ktp \nonumber 
    \\
    & \triangleq \iint_\rpos \iint_{\rposp} \spatial{K}^\dagger(\rpos) \cdot \mat{\spatialscal{C}}_\text{abs}^K(\rpos, \rposp) \cdot \spatial{K}(\rposp) \, d\rpos \, d\rposp 
\end{align}
with 
\begin{align}
\label{eq:appendix09}
    \mat{\spatialscal{C}}_\text{abs}^K(\rpos, \rposp)
    & =
    \dfrac{\eta_0^2 k_0^2}{4} \iint_{\kt} \iint_{\ktp}
    \dfrac{1}{(k_z^+)^* {k'}_z^+} \mat{\spectralscal{d}}^*(\kt) 
    \cdot \mat{\spectralscal{C}}_\text{abs}(\kt, \ktp)
    \cdot \mat{\spectralscal{d}}^T(\ktp)
     \\
    & \hspace{2cm} 
    \times \exp\Big(j (k'_x x' + k'_y y' - k_x x - k_y y)\Big) \,  d\kt \, d\ktp \nonumber
\end{align}
Note that, as defined in the Letter, $d\kt = dk_x \, dk_y /4\pi^2 $.

\subsubsection{Using current-based Kirchhoff's law}
Using the current-based mixed losses defined in Equation \eqref{eq:appendix02} and combining it with the current-based Kirchhoff's law of Equation \eqref{eq:appendix09}, one obtains: 
\begin{align}
\label{eq:appendix10}
    \mat{\spatialscal{C}}_\text{em}(\rpos, \rposp)
    & =
    \Theta(T) \eta_0^2 k_0^2 \iint_{\kt} \iint_{\ktp}
    \dfrac{1}{(k_z^+) \big({k'}_z^+\big)^*} \, \mat{S} \cdot \mat{\spectralscal{d}}(\kt)
    \cdot \mat{\spectralscal{C}}_\text{abs}^*(\kt, \ktp)
    \cdot \mat{\spectralscal{d}}^\dagger(\ktp) \cdot  \mat{S}
     \\
    & \hspace{5cm} 
    \times \exp\Big(j (k_x x + k_y y - k'_x x' - k'_y y')\Big) \,  d\kt \, d\ktp \nonumber
\end{align}

\subsubsection{Computation of the spectral content}
The spectral transform described in Equation \eqref{eq:appendix11} can be combined with the definition of the cross-spectral power density tensor in the spectral domain (Eq. (14)) to obtain the Kirchhoff's law formulated in the spectral domain:
\begin{align}
    \mat{\spectralscal{C}}_\text{em}(\kt, \ktp) 
    & \triangleq \langle \spectral{F}^\text{em}(\kt) \spectral{F}^{\text{em},\dagger}(\ktp) \rangle 
    \\
    &= \Bigg\langle \iint_\rpos \dinvspectral(\kt) \cdot \spatial{F}^\text{em}(\rpos) \exp\big( j (k_x x + k_y y) \big) \, d\rpos
    \\ \nonumber
    & \hspace{3cm}\times \bigg( \iint_{\rposp} \dinvspectral(\ktp) \cdot \spatial{F}^\text{em}(\rposp) \exp\big( j (k'_x x' + k'_y y') \big) \, d\rposp \bigg)^\dagger
    \Bigg\rangle
    \\ \label{eq:appendix18}
    &= \iint_\rpos \iint_{\rposp} \dinvspectral(\kt) \cdot 
    \langle \spatial{F}^\text{em}(\rpos) \spatial{F}^{\text{em},\dagger}(\rposp) \rangle 
    \cdot \big(\dinvspectral(\ktp)\big)^\dagger
    \\ \nonumber
    & \hspace{3cm}\times
    \exp\big(j (k_x x + k_y y - k'_x x - k'_y y)\big) \, d\rpos \, d\rposp 
\end{align}
The averaged value in \eqref{eq:appendix18} corresponds to the cross-spectral power density tensor whose value was determined in Equation \eqref{eq:appendix10}. After straightforward calculations, it leads to:
\begin{align}
    \mat{\spectralscal{C}}_\text{em}(\kt, \ktp) &= \dfrac{\Theta(T) \eta_0^2 k_0^2}{k_z^+ ({k'}_z^+)^*}  \, \mat{P} \cdot \mat{\spectralscal{C}}_\text{abs}^*(-\kt, -\ktp) \cdot \mat{P}
    \\
    & \triangleq \dfrac{\spectralscal{C}_\text{Th}}{k_z^+ ({k'}_z^+)^*} \, \mat{P} \cdot \mat{\spectralscal{C}}_\text{abs}^*(-\kt, -\ktp) \cdot \mat{P}
\end{align}
with $\spectralscal{C}_\text{Th} = \eta_0^2 \, k_0^{2} \, \Theta(T)$.


\subsection{Spatial formulation}
To prove Eq. (21), the philosophy is slightly different. Using the equivalence principle, it is easy to find a current distribution that reproduces a given incident fields. However, without knowing a-priori the shape of the surface, finding the incident fields distribution generated by a given current distribution is a more complex operation. 

In this proof, we consider that the current-based mixed losses $\mat{\spatialscal{C}}_\text{abs}^K$ are known. Then, the proof proceeds as follows:
\begin{enumerate}
    \item For a given incident fields distribution, find a currents distribution along the surface of the volume that generates the same field distribution.
    \item The power dissipated by the structure for a given incident fields distribution is equal to the power dissipated by the structure when it is illuminated by a currents distribution that generates the same incident fields. Thus, from the current-based mixed losses, build the field-based mixed losses.
    \item Using the current-based formulation of Kirchhoff's law, predict the value of the cross-spectral power density tensor along the interface.
    \item Compare both values.
\end{enumerate}

\subsubsection{Currents distribution generating a given incident fields distribution}
The incident fields are defined as the fields that would be generated by the sources located outside of the volume containing the structure if the latter was empty. Thus, the incident fields generated by a given source distribution in the presence of the structure correspond to the total fields generated by these sources without the structure. An equivalent electric ($\spatial{J}^\text{in}$) and magnetic ($\spatial{M}^\text{in}$) currents distribution on the surface generating identical fields can thus be found using the equivalence principle~\cite{Guissard_2003}. On any position $\rpos$ on the surface $\partial \Omega$ of the volume, these equivalent currents satisfy
\begin{align}
\label{eq:appendix12}
    \spatial{J}^\text{in}(\rpos) = - \spatial{n}(\rpos) \times \spatial{H}^\text{in}(\rpos) \\
    \spatial{M}^\text{in}(\rpos) = \spatial{n}(\rpos) \times \spatial{E}^\text{in}(\rpos)
\end{align}
with $\spatial{n}= (n_x; n_y; n_z)$ the outer normal to the surface and $\spatial{E}^\text{in}$ and $\spatial{H}^\text{in}$ the incident electric and magnetic fields generated by the sources. We define the tensor $\matn$ as 
\begin{equation}
    \matn(\rpos) \triangleq 
    \begin{bmatrix}
        0 & 0 & 0
            & 0 & -n_z(\rpos) & n_y(\rpos) \\
        0 & 0 & 0
            & n_z(\rpos) & 0 & -n_x(\rpos) \\
        0 & 0 & 0
            & -n_y(\rpos) & -n_x(\rpos) & 0 
        \\
         0 & -n_z(\rpos) & n_y(\rpos)
            & 0 & 0 & 0 \\
         n_z(\rpos) & 0 & -n_x(\rpos)
            & 0 & 0 & 0 \\
         -n_y(\rpos) & -n_x(\rpos) & 0
            & 0 & 0 & 0 \\
    \end{bmatrix}
\end{equation}
Defining the position-dependent column vector $\spatial{K}^\text{in} = [\spatial{J}^\text{in}; \spatial{M}^\text{in}/\eta_0]$, \eqref{eq:appendix12} can be reformulated as
\begin{equation}
\label{eq:appendix13}
    \spatial{K}^\text{in}(\rpos) =  \dfrac{1}{\eta_0} \matn(\rpos) \cdot \mat{S} \cdot \spatial{F}^\text{in}(\rpos)
\end{equation}

\subsubsection{Field-based mixed losses}
The power dissipated by the structure when illuminated by given incident fields $\spatial{F}^\text{in}$ can be estimated from the equivalent currents:
\begin{align}
    \Pabs 
    & =  \oiint_{\rposp \in \partial \Omega} \oiint_{\rpos \in \partial \Omega} \spatial{K}^{\text{in}, \dagger}(\rpos) \cdot \mat{\spatialscal{C}}_\text{abs}^K(\rpos, \rposp) \cdot \spatial{K}^\text{in}(\rpos) \, d\rpos \, d\rposp
    \\
    & = \dfrac{1}{\eta_0^2} \oiint_{\rposp \in \partial \Omega} \oiint_{\rpos \in \partial \Omega} \vect{F}^{\text{in} \dagger(\rpos)} \cdot \Big( \mat{S}^\dagger \cdot \matn^\dagger(\rpos) \cdot \mat{\spatialscal{C}}_\text{abs}^K(\rpos, \rposp) \cdot \matn(\rposp) \cdot \mat{S} \Big) \cdot \vect{F}^\text{in}(\rposp) d\rpos \, d\rposp
    \\
    & \triangleq \oiint_{\rposp \in \partial \Omega} \oiint_{\rpos \in \partial \Omega} \vect{F}^{\text{in}\dagger}(\rpos) \cdot \mat{\spatialscal{C}}_\text{abs}^\text{tan}(\rpos, \rposp) \cdot \vect{F}^\text{in}(\rposp) d\rpos \, d\rposp
\end{align}
with 
\begin{equation}
\label{eq:appendix14}
    \mat{\spatialscal{C}}_\text{abs}^\text{tan}(\rpos, \rposp) = \dfrac{-1}{\eta_0^2} \mat{S} \cdot \matn(\rpos) \cdot \mat{\spatialscal{C}}_\text{abs}^K(\rpos, \rposp) \cdot \matn(\rposp) \cdot \mat{S}
\end{equation}

\subsubsection{Cross-spectral power density tensor}
Using Equation \eqref{eq:appendix03}, the cross-spectral power density tensor of the fields emitted by the structure can be inferred from the current-based mixed losses:
\begin{equation}
\label{eq:appendix15}
    \mat{\spatialscal{C}}_\text{em}(\rpos, \rposp) = 4 \Theta(T) \mat{S} \cdot \mat{\spatialscal{C}}_\text{abs}^{K,*}(\rpos, \rposp) \cdot \mat{S}
\end{equation}

\subsubsection{Comparison of the mixed losses and cross-spectral power density tensor}
Right- and left-multiplying \eqref{eq:appendix15} by $\mat{S}$, one obtains
\begin{equation}
    \mat{\spatialscal{C}}_\text{abs}^{K}(\rpos, \rposp) = \dfrac{1}{4\Theta(T)}  \mat{S} \cdot \mat{\spatialscal{C}}_\text{em}^*(\rpos, \rposp) \cdot \mat{S}
\end{equation}
This result can be substituted into Equation \eqref{eq:appendix14}, leading to
\begin{align}
    \mat{\spatialscal{C}}_\text{abs}^\text{tan}(\rpos, \rposp) 
    &= \dfrac{-1}{4 \Theta(T) \eta_0^2} \matn(\rpos) \cdot \mat{\spatialscal{C}}_\text{em}^*(\rpos, \rposp) \cdot \matn(\rposp) 
    \\
    &= \dfrac{-1}{4 \Theta(T) \eta_0^2} \matn(\rpos) \cdot \mat{\spatialscal{C}}_\text{em}^{\text{tan},*}(\rpos, \rposp) \cdot \matn(\rposp) \label{eq:appendix16}
\end{align}
where we used the fact that $\matn \cdot \matn \cdot \matn = - \matn$.

Similarly, left- and right-multiplying equation \eqref{eq:appendix16} by $\matn$, one obtains
\begin{align}
    \mat{\spatialscal{C}}_\text{em}^{\text{tan}}(\rpos, \rposp) 
    &= - \spatialscal{C}_\text{Th} \matn(\rpos) \cdot \mat{\spatialscal{C}}_\text{abs}^{\text{tan},*}(\rpos, \rposp) \cdot \matn(\rposp)
    \\
    &= \spatialscal{C}_\text{Th} \matn(\rpos) \cdot \mat{\spatialscal{C}}_\text{abs}^{\text{tan},*}(\rpos, \rposp) \cdot \matn^T(\rposp)
\end{align}
with $\spatialscal{C}_\text{Th} = 4 \Theta(T) \eta_0^2$. 

\bibliography{sample}

%
%
%
%
%
